\documentclass[aps]{revtex4}

\usepackage{amssymb}
\usepackage[tbtags]{amsmath}
\usepackage{graphicx}
\usepackage{subfigure}

\newcommand{\SSC}{\scriptscriptstyle}
\newcommand{\double}{\baselineskip 1.24 \baselineskip}

\DeclareMathOperator{\grad}{grad}

\begin{document}
\title{Energy Differential Structure and Exchange of A
 Micro Flux Increment of Charged Particles in Longitudinal Acceleration}

\author{Ji Luo, Chuang Zhang}
\email{luoji@mail.ihep.ac.cn}
\affiliation{Accelerator Center, Institute of High Energy Physics, Beijing, China}

\date{\today}

\begin{abstract}
Differential energy structure of a micro multi-charged-particle system and the beam
internal potential energy is derived with consequent property and necessary inference.
Then by combining the energy differential structure with differential transfer
relations of physical flux density between different time domains, we present a general
explicit formula for energy exchange in the beam element's transfer or accelerating
process between source and observer, further give an analysis on relativistic beam's
mass-energy relation and beam's energy up limits for resonant frequency (rf) and DC acceleration. Finally we
discuss the property of elementary charge density field and consequences.
\end{abstract}

\maketitle

\section{Introduction}
Tracing the energy gain process of high energy electrons beam through rf acceleration,
it would be known that the longitudinally internal charge density field's affection or
contribution to the total energy of the group of electrons should be considered as the
electron number density in longitudinal axis increases progressively due to accumulated
compression, especially when its average velocity of the group of electrons becomes
almost independent to the energy after the velocity approaching light speed. However,
their functions to energy flux density basically are neglected except treating them for
beam instability and impedance\ \cite{Alexander:ape,Alexander:collective,%
Edwards:iphea,StanleyHumphries:pcpa}. From general mechanics, the
multi-particles system's energy in complete inertial frame where both
$\vec{f}(\dot{t})=m\vec{\ddot{r}}$ and $\vec{f}(\dot{t})\cdot dr=dE(\dot{t})$ exist
simultaneously can be divided into two essential parts: mass center's kinetic energy
and system's internal energy; the latter could be further divided into two
constituents: relative mass center's kinetic energy and internal potential energy whose
rate equals the rate of work done by internal forces negatively. Upon number density's
instantaneous distribution in longitudinal axis and corresponding velocity distribution
of the micro particle system, we set up the explicit formulas on mass-center kinetic
energy, relative mass center kinetic energy, internal electrical potential energy and
its changing rate. Then based on differential transfer relation of energy
flux \cite{luoji:transferrelation} and the energy differential structure, we analyze and discuss the
affection and contribution of the internal charge field to energy flux and energy
exchange as well as other consequences.
\section{Essential Interactive Manner and Energy Constituents
of Multi-Particles System}
\subsection{Interaction and Power Exchange in A complete Inertial
Frame} In a complete inertial frame \cite{luoji:frametransform}, where both
$\vec{f}_{neti}(\dot{t})=\vec{f}_{oi}(\dot{t})+\vec{f}_{ii}(\dot{t})
=m_i\vec{\ddot{r}}_{i}(\dot{t})$ and
$\vec{f}(\dot{t})\cdot\vec{\dot{r}}(\dot{t})=\frac{dE(\dot{t})}{d\dot{t}}$ exist
simultaneously, and for a multi-particle system consisted of $\Delta N$ particles, there
is the system's power exchange relation:

\begin{widetext}
\begin{equation}
\label{eq201} \sum \left[ \vec{f}_{oi}(\dot{t})+
\vec{f}_{ii}(\dot{t})\right]\cdot
\vec{\dot{r}}_{i}(\dot{t})=\vec{\dot{r}}_{mc}(\dot{t})\cdot \sum
\vec{f}_{oi}(\dot{t})+\sum
\vec{f}_{oi}(\dot{t})\cdot\vec{\dot{r}}_{imc}(\dot{t})+\sum
\vec{f}_{ii}(\dot{t})\cdot\vec{\dot{r}}_{imc}(\dot{t})
\end{equation}
\begin{equation}
\label{eq201b} \tag{\ref{eq201}$'$} \sum \left[
\vec{f}_{oi}(\dot{t})+ \vec{f}_{ii}(\dot{t})\right]\cdot
\vec{\dot{r}}_{i}(\dot{t})=\sum m_i\cdot
\vec{\ddot{r}}_i(\dot{t})\cdot\vec{\dot{r}}_{i}(\dot{t})=\frac{d}{d\dot{t}}\left[
\sum\frac{m_i}{2}
v^2_{mc}(\dot{t})\right]+\frac{d}{d\dot{t}}\left[
\sum\frac{m_i}{2} v^2_{imc}(\dot{t})\right]
\end{equation}
note: \qquad $\sum m_i\left[ \vec{r}_i(\dot{t})-\vec{r}_{mc}(\dot{t})\right]=0$; %
$\dfrac{d^n}{d\dot{t}^n} \left\{ \sum m_i \left[
\Vec{\dot{r}}_i(\dot{t})-\Vec{\dot{r}}_{mc}(\dot{t})\right]\right\}=0 \qquad n=1,2,\cdots$
\begin{equation}
\label{eq202} \Vec{\dot{r}}_{mc}(\dot{t})\cdot \sum
\Vec{f}_{oi}(\dot{t})+\sum \Vec{f}_{oi}(\dot{t})\cdot
\Vec{\dot{r}}_{imc}(\dot{t})=\frac{d}{d\dot{t}}\left[\frac{v^2_{mc}(\dot{t})}{2}\sum
m_i\right]+\frac{d}{d\dot{t}}\left[\sum
\frac{m_i}{2}v^2_{imc}(\dot{t})\right]-\sum
\Vec{f}_{ii}(\dot{t})\cdot \Vec{\dot{r}}_{imc}(\dot{t})
\end{equation}
\end{widetext}
\subsection{Energy Components of $\Delta N$ Same Charged Particles
System}

From (\ref{eq202}) and $m_i=m_q$, the power exchange relation between micro
multi-particle system and external system is:
\begin{widetext}
\begin{equation}
\label{eq202b}\tag{\ref{eq202}$'$}
\begin{split}
\frac{d\Delta E_{exc}(\dot{t})}{d\dot{t}}&=\sum%
\Vec{f}_{oi}(\dot{t})\cdot\Vec{\dot{r}}_i(\dot{t}) \\
&=\Vec{\dot{r}}_{mc}(\dot{r})\cdot\sum \Vec{f}_{oi}(\dot{t})+\sum
\Vec{f}_{oi}(\dot{t})\cdot\Vec{\dot{r}}_{imc}(\dot{t})\\
&=\frac{d}{d\dot{t}}\left[ \frac{\Delta N
m_q}{2}v_{mc}^2(\dot{t})\right]+\frac{d}{d\dot{t}}
\left[\sum\frac{m_q}{2}v_{imc}^2(\dot{t})\right]+\frac{d\Delta
E_p(\dot{t})}{d\dot{t}}\\
&=\frac{d}{d\dot{t}}\left[ \frac{\Delta N
m_q}{2}v_{mc}^2(\dot{t})\right]+\frac{d\Delta
E_i(\dot{t})}{d\dot{t}}
\end{split}
\end{equation}
\end{widetext}
here \[
\frac{d\Delta
E_p(\dot{t})}{d\dot{t}}=-\sum\Vec{f}_{ii}(\dot{t})\cdot\Vec{\dot{r}}_{imc}(\dot{t})\hspace{5.5cm}
\]

Take integral form of Eq.\ (\ref{eq202}) over $[t,t']$ or the micro multi-particle
system's whole being accelerated process
\begin{equation}
\label{eq203} \begin{split} \Delta
E_{exc}(t')&=\left.\left[\frac{\Delta N
m_q}{2}v_{mc}^2(\dot{t})+\sum
\frac{m_q}{2}v_{imc}^2(\dot{t})+\Delta
E_p(\dot{t})\right]\right|_t^{t'}\\
&=\left.\left[\frac{\Delta N m_q}{2}v_{mc}^2(\dot{t})+\Delta
E_i(\dot{t})\right]\right|_t^{t'}\\
&=\left.\Delta E(\dot{t})\right|_t^{t'}
\end{split}
\end{equation}
here:
\[
\Delta E_{exc}(t')=\int_t^{t'}\frac{d\Delta E_{exc}(\dot
t)}{d\dot{t}}d\dot{t}
\]

Furthermore, concrete formula of $\Delta E_{exc}(\dot{t})$ is
determined by energy differential structure of micro particles
system, or by explicit formulas of
$v_{mc}(\dot{t})$,\;$\sum\frac{m_q}{2}v_{imc}^2(\dot{t})$ and
$\Delta E_p(\dot{t})$ along longitudinal acceleration direction.

\section{Energy Differential Structure of Micro Multi-Charged-
Particles System}
\subsection{Kinetic Energy}
\begin{figure}
\centering \subfigure[longitudinal distribution of number density and %
velocity]{\label{fig1:a}
\includegraphics[width=8cm]{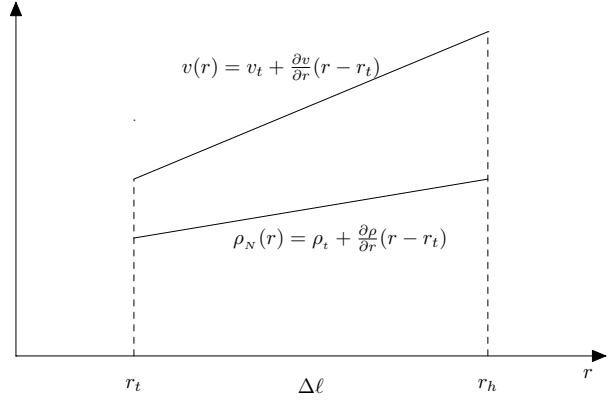}}
\\%\vspace{5mm}
\centering \subfigure[longitudinal length of the micro system]{\label{fig1:b}
\includegraphics[width=8cm]{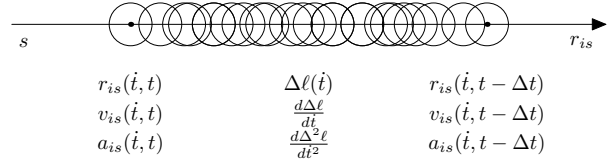}}
\caption{Diagram of micro structure along axial dimension} \label{fig1}
\end{figure}
Ref.\ Fig.\ (\ref{fig1}): longitudinal distribution of number density and corresponding velocity.
\begin{equation}
\label{eq301}
\rho_{\SSC
N}(\dot{t},r)=\rho_t(\dot{t})+\frac{\partial\rho_{\SSC
N}}{\partial r}(r-r_t)
\end{equation}
and
\begin{equation}
\label{eq301b} \tag{\ref{eq301}$'$} \int_{r_t}^{r_h}\rho_{\SSC
N}(r)dr=\Delta \ell(\rho_t+\frac{1}{2}\frac{\partial
\rho}{\partial r}\Delta \ell)=\Delta \ell{\bar\rho}_{\SSC
N}=\Delta N
\end{equation}

\begin{equation}
\label{eq302} v(\dot{t},r)=v_t(\dot{t})+\frac{\partial v}{\partial
r}(r-r_t)
\end{equation}

From Eq.\ (\ref{eq301}) and (\ref{eq301b}), there
\[
\rho_m(\dot{t},r)=m_q\rho_{\SSC
N}(\dot{t},r)
\]
\begin{equation}
\label{eq303} \int_{r_t}^{r_h}\rho_m(r)dr=m_q\Delta \ell
\bar{\rho}_{\SSC N}=\Delta \ell \bar{\rho}_m=m_q\Delta N=\Delta M
\end{equation}
\[
\rho_q(\dot{t},r)=q\rho_{\SSC N}(\dot{t},r)
\]
\begin{equation}
\label{eq304} \int_{r_t}^{r_h}\rho_q(r)dr=q\Delta \ell
\bar{\rho}_{\SSC N}=\Delta \ell \bar{\rho}_q=q\Delta N=\Delta Q
\end{equation}

For $\int_{r_t}^{r_h}(r-r_{mc})dm(r)=0$
\begin{equation}
\label{eq305}r_{mc}-r_t-\frac{\Delta
\ell}{2}=r_{mc}-r_h+\frac{\Delta \ell}{2}=\frac{\Delta
\ell^3}{12\Delta N}\frac{\partial \rho_{\SSC N}}{\partial r}
\end{equation}

\begin{widetext}
\begin{equation}
\label{eq306} v_{mc}-v_t-\frac{\Delta
\dot\ell}{2}=v_{mc}-v_h+\frac{\Delta \dot\ell}{2}=\frac{\Delta
\ell^2}{4\Delta N}\frac{\partial \rho_{\SSC N}}{\partial r}\Delta
\dot\ell+\frac{\Delta \ell^3}{12\Delta
N}\frac{\partial^2\rho_{\SSC N}}{\partial\dot t\partial r}
\end{equation}
and for $\int_{r_t}^{r_h}\left[ v(r)-v_{mc}(r_{mc})\right]dm(r)=0$ to reduce Eq.\
(\ref{eq306})
\begin{equation}
\label{eq306b} \tag{\ref{eq306}$'$} v_{mc}-v_t-\frac{\Delta
\dot\ell}{2}=v_{mc}-v_h+\frac{\Delta \dot\ell}{2}=\frac{\Delta
\dot\ell}{12\Delta N}\frac{\partial \rho_{\SSC N}}{\partial
r}\Delta \ell^2=\frac{\rho_h-\rho_t}{12\bar{\rho}_{\SSC N}}\Delta
\dot\ell
\end{equation}
thus
\begin{equation}
\label{eq307} \frac{\Delta N m_q}{2}v_{mc}^2=\frac{\Delta
M}{2}\left(v_h-\frac{\Delta \dot\ell}{2}+\frac{\Delta \dot\ell
\Delta \ell^2}{12\Delta N}\frac{\partial \rho_{\SSC N}}{\partial
r}\right)^2
\end{equation}

\begin{equation}
\label{eq308}
\begin{split}
 \sum\frac{m_q}{2}v_{imc}^2(\dot
t)&=\int_{r_t}^{r_h}dm(r)\left[v(r)-v_{mc}\right]^2\\
&=\frac{m_q}{2}\left[\Delta
N\left(v_{mc}-v_t\right)^2-(v_{mc}-v_t)(\Delta
N+\frac{1}{6}\frac{\partial \rho_{\SSC N}}{\partial r}\Delta
\ell^2+\frac{1}{3}(\Delta N+\frac{1}{4}\frac{\partial \rho_{\SSC
N}}{\partial r}\Delta \ell^2)\Delta \dot\ell^2\right]\\
&=\frac{\Delta M}{24}\Delta \dot\ell^2\left[
1-\frac{1}{12\bar\rho_{\SSC N}^2}\left(\frac{\partial \rho_{\SSC
N}}{\partial r}\Delta \ell\right)^2\right]
\end{split}
\end{equation}
\end{widetext}

Eq.\ (\ref{eq307}) and (\ref{eq308}) is kinetic energy formulas that contain the
information of the micro particles system. Eq.\ (\ref{eq308}) is kinetic energy spread
of the micro system.

\subsection{Internal Potential Energy and Its Property}
Along axis we assume average repulsive charge field force is
proportion to the difference of compressed charge density and
uncompressed charge density, or critical charge density
$\rho_{q0}$; beam transverse radius keep in constant.

\begin{eqnarray}
\label{eq309} \bar f_q(\Delta \ell)&=&\mu [\bar\rho_q(\dot
t)-\rho_{q_0}]=\mu q[\bar\rho_{\SSC N}(\dot t)-\rho_{\SSC
N_0}]\\
&=&\mu q\Delta N\left[\frac{1}{\Delta \ell(\dot
t)}-\frac{1}{\Delta \ell_0}\right]
\end{eqnarray}
where\\
$\rho_{q_0}=q\rho_{\SSC N_0}=q\frac{\Delta N}{\Delta \ell_0}$\\
$\mu$ --- compression coefficient of charge density $[\mu]=\left[\frac{N
m^3}{C}\right]$\\
$
\bar f_q(\Delta \ell)=\begin{cases}\mu q\Delta
N(\frac{1}{\Delta \ell}-\frac{1}{\Delta \ell_0})&\quad\Delta \ell < \Delta \ell_0\\
0&\quad\Delta \ell \geqslant \Delta \ell_0\end{cases}
$

\begin{equation}
\label{eq310}
\begin{split}
\Delta E_p[\Delta \ell(\dot t)]&=\int_{\Delta \ell}^{\Delta \ell_0}\!\!\!\bar f(\Delta
\ell)d\Delta \ell=\int_{\Delta \ell}^{\Delta \ell_0}\!\!\!\!\mu q\Delta
N(\frac{1}{x}-\frac{1}{\Delta \ell_0})dx\\
&=\mu q\Delta N\left[\ln\frac{\Delta \ell_0}{\Delta \ell(\dot
t)}-\frac{\Delta \ell_0-\Delta \ell(\dot t)}{\Delta \ell_0}\right] \\
&=\mu q\Delta N(\ln\frac{\Delta \ell_0}{\Delta \ell}+\frac{\Delta \ell}{\Delta
\ell_0}-1)\\
&=\!\!\!\!\sum^{\SSC{\Delta N=\sum n_i}}\!\!\!\!n_i \mu q\left[
\ln\frac{\lambda_0}{\lambda_i(\dot t)}+\frac{\lambda_i(\dot t)}{\lambda_0}-1\right]
\end{split}
\end{equation}
$\lambda_0$--- individual charge field's critical diameter; $\lambda$--- compressed
diameter of individual charge field; $\lambda\in(0,\lambda_0]$.
here define
\[
k(\dot t)=\frac{\lambda(\dot t)}{\lambda_0}=e^{-\eta(\dot t)}
\]
\double
\begin{center}
$k(\dot t)\in(0,1]$,\; $\eta(\dot t)\in[0,+\infty)$\\
\double $k(\dot t)$--- ratio of compressed and intrinsic volumes of individual charge
field.\\
$\eta(\dot t)$--- potential energy index of individual charge
field.\\
\end{center}
and
\begin{equation}
\label{eq311} \frac{d\Delta E_p[\Delta \ell(\dot t)]}{d\dot
t}=\frac{d\Delta E_p}{d\Delta \ell}\Delta \dot\ell=-\mu
q\rho_{\SSC N_0}\left(\frac{\Delta \ell_0}{\Delta
\ell}-1\right)\Delta \dot\ell
\end{equation}

Further
\begin{equation}
\label{eq312} \begin{split}\sum f_{ii}(\dot t)\cdot\dot
r_{imc}(\dot
t)&=\int_{r_t}^{r_h}f_i(r)\frac{\partial v}{\partial r}dr\\
&=\int_{r_t}^{r_h}\mu q[\rho_{\SSC N}(r)-\rho_{\SSC N_0}]\frac{\partial v}{\partial r}dr\\
&=\mu q\rho_{\SSC N_0}\left(\frac{\Delta \ell_0}{\Delta
\ell}-1\right)\Delta \dot\ell
\end{split}
\end{equation}

Compare Eq.\ (\ref{eq311}) and (\ref{eq312})

\begin{equation}
\label{eq312b}\tag{\ref{eq312}$'$}\frac{d\Delta E_p(\dot t)}{d\dot
t}=-\int_{r_t}^{r_h}f_i(r)\frac{\partial v}{\partial r}dr
\end{equation}

From Eq.\ (\ref{eq202b}) suppose $\frac{d\Delta E_{exc}(\dot t)}{d\dot t}=0$; there
$\frac{d\Delta E_i(\dot t)}{d\dot t}=0$, so

\begin{equation}
\label{eq313}\frac{d\Delta E_p(\dot t)}{d\dot t}=-\frac{d}{d\dot
t}\left[\sum\frac{m_q}{2}v_{imc}^2(\dot t)\right]
\end{equation}

\begin{equation}
\label{eq313b}
\tag{\ref{eq313}$'$}\text{or}\qquad\int_{r_t}^{r_h}f_i(r)\frac{\partial
v}{\partial r}dr=\frac{d}{d\dot
t}\left[\sum\frac{m_q}{2}v_{imc}^2(\dot t)\right]
\end{equation}

It is shown in Eq.\ (\ref{eq313}), (\ref{eq313b}) that for internal force to do
positive work will elongate the length, enlarge kinetic energy spread and exhaust or
consume the internal potential energy; for the force to do negative work, the length
will be compressed and kinetic energy spread reduced, internal potential energy
increased. However once the internal force starts to do negative work, the process will
be irreversible without external power's joining.
\subsection{Energy Differential Structure of $\Delta E(\dot t)$
Carried by the Micro Multi-Particle System}
\begin{widetext}
\begin{equation}
\label{eq314}
\begin{split}
\Delta E(\dot t)&=\frac{\Delta N m_q}{2}v_{mc}^{2}(\dot
t)+\int_{r_t}^{r_h}\frac{dm(r)}{2}v_{imc}^{2}(r)+\Delta E_p(\dot
t)\\
&=\frac{\Delta M}{2}\left(v_h-\frac{\Delta
\dot\ell}{2}+\frac{\Delta \dot\ell}{12\Delta N}\frac{\partial
\rho_{\SSC N}}{\partial r}\Delta \ell^2\right)^2+\frac{\Delta
M}{24}\Delta \dot\ell^2\left[1-\frac{1}{12\bar\rho_{\SSC
N}^2}\left(\frac{\partial
\rho_{\SSC N}}{\partial r}\Delta \ell\right)^2\right]+\mu \Delta Q\left(\ln\frac{\Delta \ell_0}{\Delta
\ell}+\frac{\Delta \ell}{\Delta \ell_0}-1\right)\\
&=\frac{\Delta M}{2}\left(v_h-\frac{\Delta
\dot\ell}{2}+\frac{\Delta \dot\ell}{12\bar\rho_{\SSC
N}}\frac{\partial \rho_{\SSC N}}{\partial r}\Delta
\ell\right)^2+\frac{\Delta M}{24}\Delta
\dot\ell^2\left[1-\frac{1}{12\bar\rho_{\SSC
N}^2}\left(\frac{\partial
\rho_{\SSC N}}{\partial r}\Delta \ell\right)^2\right]+\!\!\!\!\sum^{\SSC{\Delta N=\sum n_i}}\!\!\!\!n_i \mu q\left[
\ln\frac{\lambda_0}{\lambda_i(\dot t)}+\frac{\lambda_i(\dot t)}{\lambda_0}-1\right]
\end{split}
\end{equation}
\end{widetext}

Eq.\ (\ref{eq314}) indicates that both kinetic and potential energy have possible
distributions; for kinetic energy it depends on length, length's rate and $\rho_{\SSC
N}$ distribution or different compression states due to internal charge field force in
longitudinal dimension.

\section{Energy Flux and Energy Exchange}
\subsection{Energy flux density and energy flux}

At instant $\dot t\in[t,t']$ there is energy flux density
$\rho_{\SSC \Delta E}[r_{is}(\dot t)]$ on a crossing section
located at $r_{is}(\dot t)$
\begin{equation}
\label{eq401}\rho_{\SSC \Delta E}[r_{is}(\dot t)]=\rho_{\SSC
N}[r_{is}(\dot t)]\left\{\frac{m_q}{2}v_{mc}^2(\dot t)+\mu
q\left[\ln\frac{\lambda_0}{\lambda(\dot t)}+\frac{\lambda(\dot
t)}{\lambda_0}-1\right]\right\}
\end{equation}
and energy flux
\begin{widetext}
\begin{equation}
\label{eq402}
\begin{split}
\mathcal J_{\Delta E}(\dot t)&=\rho_{\SSC \Delta E}[r_{is}(\dot
t)]v_{is}(\dot t)\\
&=\rho_{\SSC N}[r_{is}(\dot t)]\left\{\frac{m_q}{2}v_{mc}^2(\dot
t)+\mu q\left[ \ln\frac{\lambda_0}{\lambda_i(\dot
t)}+\frac{\lambda_i(\dot
t)}{\lambda_0}-1\right]\right\}v_{is}(\dot t)\\
&=J[r_{is}(\dot t)]\left\{\frac{m_q}{2}v_{mc}^2(\dot t)+\mu
q\left[ \ln\frac{\lambda_0}{\lambda_i(\dot
t)}+\frac{\lambda_i(\dot t)}{\lambda_0}-1\right]\right\}
\end{split}
\end{equation}
\end{widetext}

\subsection{Differential transfer relation of energy flux density between
two time domains and energy exchange}
Ref.\ \cite{luoji:transferrelation} there
\[
\Delta E_{exc}(t')=\Delta E(t')-\Delta E_{so}(t)
\]
\[
\Delta N (t')=\Delta N(t)
\]
to multiply both equations with $\frac{1}{\Delta t}$ or $\frac{1}{\Delta t'}$, then
take limit as $\Delta t\to 0, \Delta t'\to 0$
\begin{equation}
\label{eq403}\mathcal J_{exc}(t')=\mathcal J(t')-\mathcal
J_{so}(t)\frac{dt}{dt'}
\end{equation}
\begin{equation}
\label{eq404}J(t')=J(t)\frac{dt}{dt'}\quad\text{or}\quad\rho_{\SSC
N}[r_p(t')]v_{io}(t')=\rho_{\SSC
N}[r_s(t)]v_{sis}(t)\frac{dt}{dt'}
\end{equation}

\begin{widetext}
Then combine Eq.\ (\ref{eq402}) and (\ref{eq404}) with Eq.\ (\ref{eq403})
\begin{equation}
\label{eq405}
\begin{split}
\mathcal J_{exc}(t')&=J[r_s(t)]\frac{dt}{dt'}\left.\left\{\frac{m_q}{2}v_{mc}^2(\dot
t)+\mu q\left[\ln\frac{\lambda_0}{\lambda(\dot t)}+\frac{\lambda(\dot
t)}{\lambda_0}-1\right]\right\}\right|_t^{t'}\\
&=\rho_{\SSC
N}[r_s(t)]v_{sis}(t)\left[1-\frac{dT(t')}{dt'}\right]\left.\left\{\frac{m_q}{2}v_{mc}^2(\dot
t)+\mu q\left[\ln\frac{\lambda_0}{\lambda(\dot
t)}+\frac{\lambda(\dot
t)}{\lambda_0}-1\right]\right\}\right|_t^{t'}\\
&=\rho_{\SSC
N}[r_o(t')]v_{io}(t')\left.\left\{\frac{m_q}{2}v_{mc}^2(\dot
t)+\mu q\left[\ln\frac{\lambda_0}{\lambda(\dot
t)}+\frac{\lambda(\dot
t)}{\lambda_0}-1\right]\right\}\right|_t^{t'}\\
&=J[r_o(t')]\left.\left\{\frac{m_q}{2}v_{mc}^2(\dot t)+\mu
q\left[\ln\frac{\lambda_0}{\lambda(\dot t)}+\frac{\lambda(\dot
t)}{\lambda_0}-1\right]\right\}\right|_t^{t'}
\end{split}
\end{equation}

By using $\lambda(t')+\lambda_c(t')=\lambda(t)$ \cite{luoji:transferrelation}, $
\lambda_c(t')=\int_t^{t'}\frac{d\lambda_c(\dot t)}{d\dot t}d\dot
t=\int_0^{\lambda_c(t')}d\lambda_c(\dot t)
\begin{cases}
>0&\quad\text{compression}\\
<0&\quad\text{elongation}
\end{cases}
$

\begin{equation}
\label{eq406}
\begin{split}\mathcal J_{exc}(t')&=\rho_{\SSC
N}[r_s(t)]v_{sis}(t)\frac{dt}{dt'}\left\{\frac{m_q}{2}\left[v_{io}^2(t')-
v_{sis}^2(t)\right]+\mu
q\left[\ln\frac{\lambda(t)}{\lambda(t')}+\frac{\lambda(t')
-\lambda(t)}{\lambda_0}\right]\right\}\\
&=J[r_s(t)]\frac{dt}{dt'}\Biggl\{\frac{m_q}{2}\left[v_{io}^2(t')-
v_{sis}^2(t)\right]+\mu
q\left\{\ln\left[1+\frac{\lambda(t)}{\lambda(t')}\right]-\frac{\lambda_c(t')
}{\lambda_0}\right\}\Biggr\}\\
&=J[r_o(t')]\Biggl\{\frac{m_q}{2}\left[v_{io}^2(t')- v_{sis}^2(t)\right]+\mu
q\left\{\ln\left[1+\frac{\lambda(t)}{\lambda(t')}\right]-\frac{\lambda_c(t')
}{\lambda_0}\right\}\Biggr\}\\
&=\rho_{\SSC N}[r_o(t')]v_{io}(t')\Biggl\{\frac{m_q}{2}\left[v_{io}^2(t')-
v_{sis}^2(t)\right]+\mu
q\left\{\ln\left[1+\frac{\lambda(t)}{\lambda(t')}\right]-\frac{\lambda_c(t')
}{\lambda_0}\right\}\Biggr\}
\end{split}
\end{equation}
\end{widetext}

Above Eq.\ (\ref{eq405}) and (\ref{eq406}) is energy exchange relation between a micro
beam element and external system during the accelerating process within which the micro
beam element travel through the space between source and detector.

For $\mathcal J_{exc}(t')>0$ external system supplies energy to beam for contributing
to either kinetic or potential energy or both as well as time affection factor.For
$\mathcal J_{exc}(t')<0$ external system gains energy from beam system, may from either
of kinetic or potential energy; in case the time factor determined by dynamical process
can also affect energy's output of micro beam element.

\subsection{Illustration or illustrative examples}
\subsubsection{Selection of time domains}
This is completely dependent to which beam's acceleration or transportation process you
are interested in. The flux source and observing location could be selected at entrance
and exit point or cross section of a resonant cavity respectively where signal's
transfer time $T(t)=T(t')$ is just charged particle's transit time. The flux source and
observing location could also be selected at beam source (gun) and beam extracting
point for tracing the whole process of the beam's acceleration and transportation drift
where signal's transfer time $T(t)=T(t')$ can be consisted by various constituent
$T_i(t_{i-1})=T_i(t_i)$ and there is consequent relation $T(t)=T(t')=\sum
T_i(t_{i-1})$ \cite{luoji:timefunction}. Once the time domains are decided, then all differential
transfer relations \cite{luoji:transferrelation} between the time domains and energy differential
information could be used to search exact solution.

\subsubsection{Energy exchange of relativistic electrons beam and
inference} In this case, $v_{io}(t')\simeq v_{sis}(t)\simeq c$ , and from Eq.\ %
(\ref{eq406}):
\begin{widetext}
\begin{equation}
\label{eq407}\begin{split} \mathcal J_{exc}(t')&\simeq \rho_{\SSC
N}[r_s(t)]c\frac{dt}{dt'}\mu
q\left\{\ln\left[1+\frac{\lambda_c(t')}{\lambda(t')}\right]-%
\frac{\lambda_c(t')}{\lambda_0}\right\}\\
&=c\mu q\rho_{\SSC
N}[r_s(t)]\frac{dt}{dt'}\left[\ln\frac{\lambda(t)}%
{\lambda(t')}-\frac{\lambda(t)-\lambda(t')}{\lambda_0}\right]\\
&=c\mu q\rho_{\SSC N}[r_s(t)]\left[1-\frac{dT(t')}{dt'}\right]\left[\ln\frac{\lambda(t)}%
{\lambda(t')}-\frac{\lambda(t)-\lambda(t')}{\lambda_0}\right]\\
&=c\mu q\rho_{\SSC N}[r_s(t)]\left[1-\frac{dT(t')}{dt'}\right]
\left\{\eta(t')+e^{-\eta(t')}-\left[\eta(t)+e^{-\eta(t)}\right]\right\}\\
&=c\mu q\rho_{\SSC N}[r_s(t)]\left[1-\frac{dT(t')}{dt'}\right]\left\{\ln\frac{k(t)}%
{k(t')}-\left[k(t)-k(t')\right]\right\}
\end{split}
\end{equation}

Thus there are consequent inferences below for constant state,
elongation's and compression's.
\begin{align*}
\intertext{A. Constant State}&\quad\frac{dT(t')}{dt'}=0,&\frac{dt}{dt'}=1,&\quad\ln\frac{\lambda(t)}{\lambda(t')}-%
\frac{\lambda(t)-\lambda(t')}{\lambda_0}=0;&\mathcal
J_{exc}(t')=0&\quad\eta(t')=\eta(t)&k(t')=k(t)\\
%\intertext{\hfill Constant State}
\intertext{B. Elongation State}&\quad\frac{dT(t')}{dt'}>0,&0<\frac{dt}{dt'}<1,&\quad\ln\frac{\lambda(t)}{\lambda(t')}-%
\frac{\lambda(t)-\lambda(t')}{\lambda_0}<0;&\mathcal
J_{exc}(t')<0&\quad\eta(t')<\eta(t)&k(t')>k(t)\\
%\intertext{\hfill Elongation State}
\intertext{C. Compression State}&\quad\frac{dT(t')}{dt'}<0,&\frac{dt}{dt'}>1,&\quad\ln\frac{\lambda(t)}{\lambda(t')}-%
\frac{\lambda(t)-\lambda(t')}{\lambda_0}>0;&\mathcal
J_{exc}(t')>0&\quad\eta(t')>\eta(t)&k(t')<k(t)\\
%\intertext{\hfill Compression State}
\end{align*}
\end{widetext}

Above inferences show that in constant state A, $\rho_{\SSC N}(t')=\rho_{\SSC N}(t)$
and without energy exchange; in elongation state B, $\rho_{\SSC N}(t')<\rho_{\SSC
N}(t)$ and internal potential energy releases to external system; in compression state
C, $\rho_{\SSC N}(t')>\rho_{\SSC N}(t)$ and internal potential energy increases result
from beam's gaining energy from external system.

Further more, we can make inferences below:

\renewcommand{\labelenumi}{(\roman{enumi})}
\begin{enumerate}
\item For relativistic electron or ion beam, it's energy increment
or increase basically results from the gain of itself's potential
energy through longitudinal compression, otherwise the beam's
energy will meet its limit, that is its kinetic energy limit.

This is main difference of rf and DC acceleration method on electron beam's energy up
limits, for rf method the bunch length can be compressed progressively further and
further so that its energy will not meet up limit; whereas using DC acceleration the
length of a micro beam element is longer than the critical length within which the
beam's internal potential energy can be existed or accumulated through further
compression, in addition, DC method has no compression mechanism. Thus DC acceleration
beam's energy has its energy limit, its kinetic limit, even DC acceleration
voltage has no technical limit.

Ref.\ \cite{luoji:transferrelation}:
%\[
\begin{eqnarray*}
&&\frac{d\ell(t')}{d\ell(t)}=\frac{v_{io}(t')}{v_{sis}(t)}\frac{dt'}{dt}%
=\frac{c}{v_{sis}(t)}\\
&&=1+\frac{1}{v_{sis}(t)}\int_t^{t'}\frac{\partial
v_{is}(\dot t,r_{is})}{\partial r_{is}}v_{is}(\dot
t,t)\frac{dt_j(\dot t)}{dt}d\dot t \gg 1
\end{eqnarray*}
%\]

We know that in DC acceleration the longitudinal length of a micro beam element is
progressively elongated.

\item The mass of relativistic beam, collectives' or individual's, is independent to
its energy.

From Eq.\ (\ref{eq407}) it is known that the energy exchange of relativistic beam
essentially depends upon beam internal potential energy's exchange with external
system, the density of the particles number at observing location will fluctuate as the
energy exchanger occur. And the number density's change is synchronized with potential
energy gain or loss of the beam, as beam's potential energy increases the density
increases, while the energy losses the density decreases. Since $\rho_{\SSC
N}[r_0(t')]=\rho_{\SSC N}[r_s(t)]\frac{dt}{dt'}=\rho_{\SSC N}[1-\frac{dT(t')}{dt'}]$
then it is clear that the density's fluctuation is time function related due to the
variation of signal transfer time.

Ref.\ \cite{luoji:transferrelation} mass flux density transfer relation
$J_m(t')+J_{loss}(t')=J_m(t)\frac{dt}{dt'}$ or
$\rho_m(t')c=\rho_m(t)c\frac{dt}{dt'}$, here $J_{loss}(t')=\frac{dm_{loss}}{dt'}=0$.

differential relation on number of particles, mass conversation
\begin{equation}
\label{eq408} \Delta N(t')+\Delta N_{loss}(t')=\Delta N(t)
\end{equation}
multiply Eq.\ (\ref{eq408}) with $m_q$:
\[
m_q\Delta N(t')+m_q\Delta N_{loss}(t')=\Delta
N(t)m_q
\]
$\Delta M(t')+\Delta M_{loss}(t')=\Delta M(t)$
here
$\Delta M_{loss}=\int_t^{t'}\frac{d\Delta
M_{loss}(\dot t)}{d\dot t}d\dot t\geqslant 0
$
there only
\begin{gather*}
\Delta M(t')\leqslant\Delta M(t)\\
c\rho_m(t')dt'\leqslant c\rho_m(t)dt
\end{gather*}

However in rf acceleration the corresponding energy increment
$\Delta E$ and energy flux $\mathcal J$
\begin{gather*}
\Delta E(t')=\Delta E(t)+\Delta E_{exc}(t')-\Delta E_c(t')\gg
\Delta E(t)\\
\mathcal J(t')dt'\gg \mathcal J(t)dt
\end{gather*}

\end{enumerate}

\section{Discussion}
\subsection{Charge field property}
Charge is structure density field possessed; for a elementary
charge $q$ there $\iiint\limits_V\rho(r)\,dV=q$ and in
uncompressed state or its intrinsic state
\[
\rho(r)=\begin{cases}\frac{2q(R_0-r)}{R^24\pi r^2}&0<r\leqslant
R_0\\0&r>R_0\end{cases}
\]
and with$\grad\rho(r)=\frac{q}{2\pi R_0^2}\frac{r-2R_0}{r^3}$
apply to $\iiint(r-r_{cc})\,dq=0$, there charge center $r_{cc}(\dot
t)$, $v_{cc}(\dot t)$ etc.
\[
\iiint\limits_V r\rho(r)\,dV=r_{cc}\iiint\limits_V\rho(r)\,dV=qr_{cc}
\]

Once charge's electromagnetic interaction and charge's potential
energy are concerned its density field effect can not be neglected
anymore while merely regarding charge as a point charge whose
charge all be centralized on its charge center for approximatively
simplified application.

Charge density field can be compressed. Based on this, its internal potential energy
can exist. In addition the compression can bring about some electromagnetic
consequences in magnetic dipole \cite{luoji:lorentz}. Density's change will produce internal
force within charge density field. The force depends upon both density's difference
between compressed density and corresponding intrinsic density, and ratio of
compression which is the intrinsic density related. So calculation of stored potential
energy of charge density field precisely can only be done through corresponding
relation $\rho_i(\dot t)\,dv_i(\dot t)=\rho_{oi}\,dv_{oi}$, this may be uneasy to
resolve since the density's spatial distribution
at any instant is not linear.

Fortunately in multi-particle system, a relative macro system, the density's
distribution at any instant comply to linear distribution as we select the longitudinal
length of the micro element short enough so that the compression ratio or coefficient
$\mu$ can be treated as constant.

In addition, for the micro beam element, or $\Delta N$ number of particles, within a
cylindrical volume with constant radius~$a$, there
\[
\frac{\Delta N}{\pi a^2}=\Delta
\ell(\dot t)\bar \rho(\dot t)=constant
\]
then
$\Delta \ell(\dot t)=\frac{\Delta N}{\pi
a^2}\lambda(\dot t)$ and $\Delta \ell_0=\frac{\Delta N}{\pi a^2}\lambda_0$,
thus
\[
\frac{\Delta \ell_0}{\Delta \ell(\dot
t)}=\frac{\lambda_0}{\lambda(\dot t)}
\]

So we can convert the ratio of the micro beam's lengthes to ratio
of individual particles'. Furthermore, we define $k$ and $\eta$,
$k(\dot t)=e^{-\eta(\dot t)}$, for indicating individual charge
field's compression and potential energy status.

\subsection{Time function related factor and energy differential
structure's implication}

\renewcommand{\labelenumi}{\Alph{enumi}}
\begin{enumerate}
\item For charged particle's longitudinal interaction, a particle behind with higher
velocity than front particle's can not pass over the ahead particle; therefore related
time function is positive reversible \cite{luoji:timefunction}, $\frac{dt'}{dt}>0$,
$\frac{dt}{dt'}>0$, or $\frac{dT(t)}{dt}>-1$ and $\frac{dT(t')}{dt'}<1$. Besides,
$\frac{dt}{dt'}$ factor can contribute to enhance or decrease output number flux, mass
flux, charge flux, energy flux $\mathcal J_{exc}(t')$ (Ref.\ Eq.\ (\ref{eq406})),
however its value is determined by energy exchange of external system and micro beam
element. \item The energy differential structure and relevant inference emphasize
beam's itself property and nature. The work, combined with differential transfer
relations \cite{luoji:transferrelation} and other tools, can help us insight beams inner behavior and
response and tracing a specified micro beam element whole physics process of beams'
energy components and exchange. With it we may evaluate some conventional dictations
and interpretations in differential angle.
\end{enumerate}

\section{Conclusion}
Energy of a micro beam element is consisted of three parts, mass center and relative
center mass kinetic energy and beam's internal potential energy. Each of them is
particle's longitudinal velocity distribution and density (number, consequent mass and
charge as well as internal charge field force) distribution related within the micro
element. The internal potential energy exists until its length shorter than the
critical length or individual charge density field's diameter shorter than its
intrinsic diameter results form accumulated compression of the beam length. And
quantitative the potential energy comply with logarithm function with respect to the
longitudinal length variable. For relativistic beam, its energy exchange essentially
occur actually between external system's energy and beam's internal potential energy;
at same time beam's mass, collective's or individual's, is independent to its energy.
Time function factor can affect various fluxes of the beam simultaneously and the
factor is determined by designated dynamical process or energy exchange between
external system and beam itself.

\end{document}